\documentclass[proceedings]{JHEP3}

\PrHEP{PrHEP hep2001}                   
\conference{International Europhysics Conference on HEP}                

\usepackage{epsfig}                   

\def\gev{{\rm GeV}}
\def\mev{{\rm MeV}}
\def\lqcd{\Lambda_{\rm QCD}}
\def\OMIT#1{{}}

\def\shat{\hat s}

\def\qsh{\hat q^2}
\def\qcut{q_{\rm cut}}
\def\mxcut{m_{\rm cut}}
\def\mxsqcut{\mxcut^2}

\def\gcut{G(\qcut^2,\mxcut)}

\def\mbups{m_b^{1S}}

\def\vereq#1#2{\lower3pt\vbox{\baselineskip1pt\lineskip1pt
     \ialign{\\$#1\hfill##\hfil\\$\crcr#2\crcr\sim\crcr}}}

\def\epsblm{\epsilon^2_{\rm BLM}}

\title{Precision determination of $V_{\rm ub}$}

\author{\speaker{Christian W.~Bauer}$^1$,                       
       Zoltan Ligeti$^2$, Michael Luke$^3$\\  
       \begin{center}
	$^1$Physics Department, University of California at San
		Diego, La Jolla, CA 92093\\ 
       	$^2$Ernest Orlando Lawrence Berkeley National Laboratory\\
		University of California, Berkeley, CA 94720 \\
      	$^3$Department of Physics, University of Toronto, 
   	 	60 St.\ George Street,\\ Toronto, Ontario, Canada M5S 1A7\end{center}          }                              

\abstract{We review how to determine $|V_{ub}|$ from inclusive semileptonic $B$ decay using
combined cuts on the leptonic and hadronic invariant masses to eliminate the
$b\to c$ background.  This leads to a determination of $|V_{ub}|$ with theoretical uncertainty at the 5--10\% level.\\
LBNL-49079, UCSD/PTH 01-22, UTPT-01-13
}

\begin{document}

The magnitude of $V_{ub}$ determines the length of one of the side of the unitarity triangle and is therefore of great importance to overconstrain this triangle. The inclusive decay rate $B \to X_u \ell \bar \nu$ is directly proportional to $|V_{ub}|^2$ and can be calculated reliably and with small uncertainties using the operators product expansion (OPE). Unfortunately, the $\sim$100 times background from $B \to X_c \ell \bar \nu$ makes the measurement of the totally inclusive an almost impossible task\footnote{It has recently been suggested that a measurement of the totally inclusive rate might be possible using totally reconstructed $B$ decays}. Several cuts have been proposed in order to reject the $b \to c$ background, however care has to be taken to ensure that the decay rate in the restricted region of phase space can still be predicted reliably theoretically. The cut which is easiest to implement experimentally is on the energy of the charged lepton, requiring $E_\ell > (m_B^2-m_D^2)/2m_B$. Unfortunately, this cut restricts the remaining region of phase space too much for the OPE to still be valid. Instead, a twist expansion has to be performed \cite{shape}, and at leading order the decay rate is determined by the light cone distribution function of the $B$ meson, with subleading twist corrections suppressed by powers of $1/m_b$ \cite{BLM}. This distribution function can not be calculated perturbatively and has to be determined experimentally, for example from the photon energy spectrum in $B \to X_s \gamma$ \cite{leptoncut}. The same light cone distribution function also determines the rate in the presence of a cut on the hadronic invariant mass \cite{sHcut}, which has the advantage of keeping most of the $b \to u$ events. In this talk I will show that using a cut on the leptonic invariant mass results in a regions of phase space that is free from $b \to c$ background and which can be calculated reliably in the standard OPE.

Using a cut on the leptonic invariant mass $q^2 = (p_\ell +p_\nu)^2$ to measure $|V_{ub}|^2$ was first proposed in \cite{qsqcut}, and it was shown that requiring $q^2 > m_B^2 - m_D^2$ leaves a region of phase space free from $b \to c$ events, which can be calculated using the usual OPE. The number of events surviving such a cut on $q^2$ were calculated in \cite{qsqcut} and depending on the exact value of the cut chosen, the fraction of events surviving the cut is between 10 and 20\%, with uncertainties on $|V_{ub}|$ ranging from 15\% for $q^2_{\rm cut} = m_B^2-m_D^2 = 11.6 {\rm GeV}^2$ to 25\% for $q^2_{\rm cut} = 14 {\rm GeV}^2$.

\EPSFIGURE{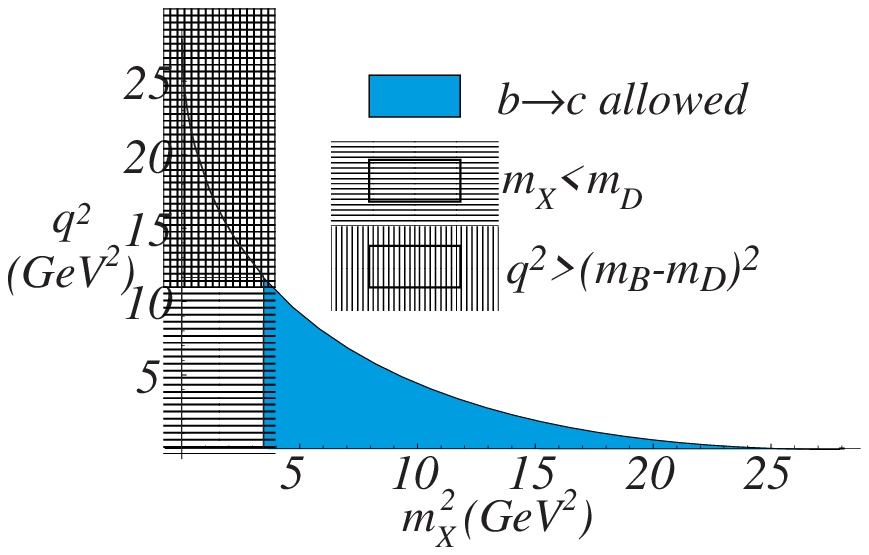,width=6cm}
{The dalitz plot for $q^2$ and $s_H$.\label{figdalitz}} 
From Fig.~\ref{figdalitz} one can see that the $q^2$ cut keeps a subset of the phase space available to a cut on the hadronic invariant mass $m_X$. Thus, a cut on $m_X$ keeps a much larger fraction of the events, up to 80\%. However, the fact that the former cut avoids the region of low $m_X$ and low $q^2$ where the structure function dominates, is why the decay rate in the presence of a $q^2$ cut can be calculated while in the presence of a pure $m_X$ cut it can not. However, it was observed in \cite{doublecut} that combining the cuts on $q^2$ and on $m_X$ allows for a much larger fraction of events compared to a pure $q^2$ cut,  while keeping the partial rate calculable using the local OPE. 

The integrated rate with a lower cut $\qcut^2$ on $q^2$ and an upper cut
$\mxcut$ on $m_X$ may be written as
\begin{equation}\label{defineg}
\int_{\hat\qcut^2}^1 {\rm d}\hat q^2 \int_0^{\hat s_0}
   {\rm d}\hat s\, {{\rm d}\Gamma\over {\rm d}\hat q^2 {\rm d}\hat s}
\equiv {G_F^2 |V_{ub}|^2\, (4.7\,\gev)^5\over 192\pi^3}\; \gcut\,,
\nonumber
\end{equation}
where where $\hat q = q/m_b$, $\shat = (v - \hat q)^2$ is the rescaled
partonic invariant mass, $v$ is the four-velocity of the decaying $B$ meson,
and
\begin{eqnarray}\label{s0limit}
\hat s_0 = \cases{
\displaystyle \left( 1 - \sqrt{\qsh}\right)^2
   & for~ $\mxcut > m_B - m_b\,\sqrt{\qsh}$\,, \vspace*{6pt}\cr
\displaystyle 0
   & for~ $\mxcut^2 < (m_B-m_b\,\qsh)\, (m_B-m_b)$\,, \vspace*{6pt}\cr
\displaystyle \frac\mxsqcut{m_B m_b} + \left( \frac{m_B}{m_b} - 1 \right)
   \left( \frac{m_b}{m_B}\,\qsh - 1 \right)
   & otherwise\,. \cr}
\end{eqnarray}
The hadronic invariant mass $m_X$ is related to $\hat q^2$ and $\hat s$ by
\begin{equation}\label{rel}
  m_X^2 = \shat\, m_B m_b  + (m_B-m_b)(m_B- \qsh m_b) \,.
\end{equation}
$\gcut$ is the ratio of the semileptonic $b\to u$ width with cuts on $q^2$ and
$m_X$ to the full width at tree level with $m_b=4.7\,\gev$.  The fraction of
semileptonic $b\to u$ events included in the cut rate is $\simeq 1.21\, \gcut$.
Note that  the
$m_b^5$ prefactor, a large source of uncertainty, is included in $\gcut$. The
theoretical uncertainty in $|V_{ub}|$ is therefore  half the
uncertainty in the prediction for $\gcut$.

\EPSFIGURE{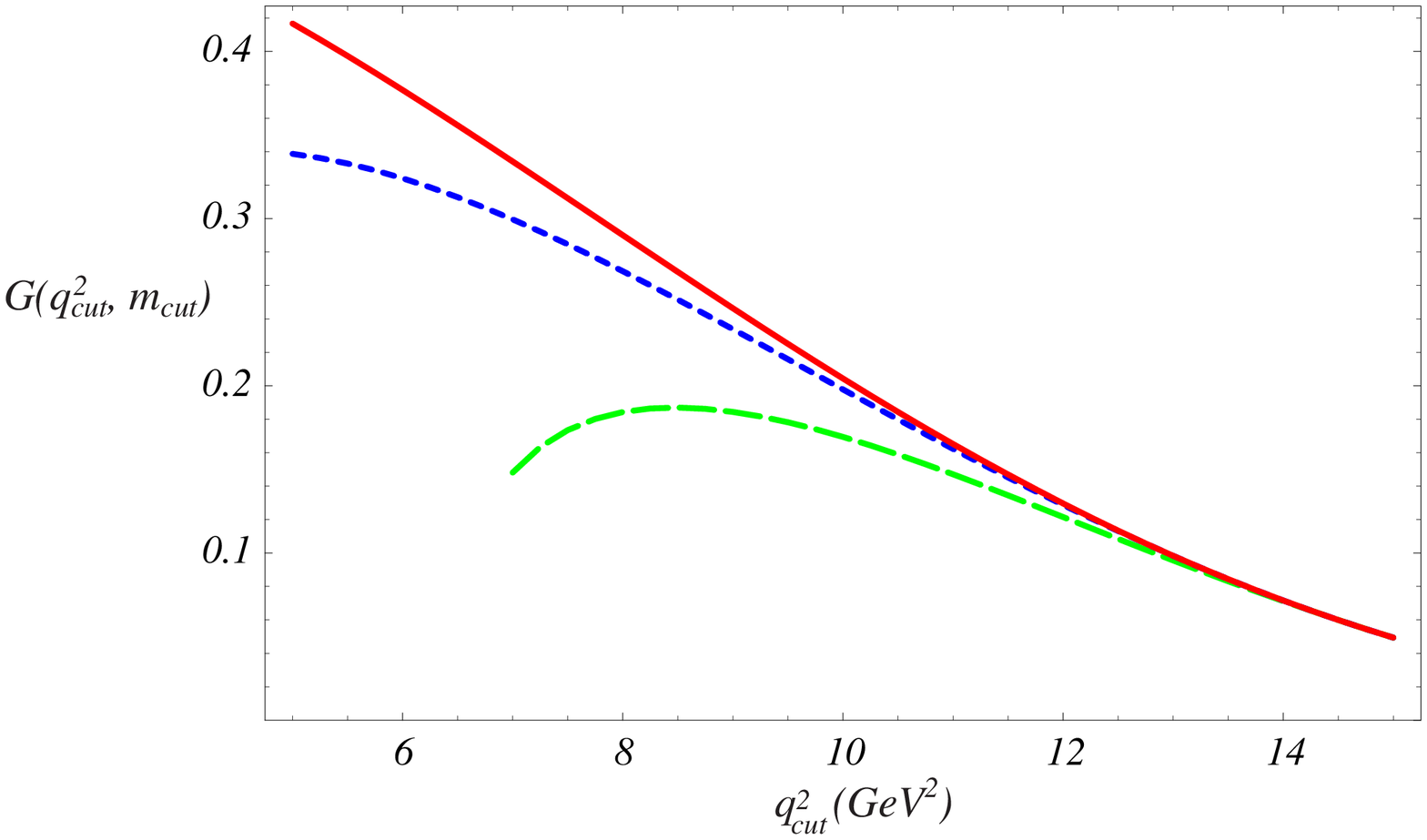,width=6cm}
{$\gcut$ as a function of 
$\qcut^2$, for $\mxcut=1.86\,\gev$ (solid  line),
$1.7\,\gev$ (short dashed line) and $1.5\,\gev$ (long dashed line). \label{figfinalplot}} 
In Ref.~\cite{doublecut} this function $\gcut$ was calculated up to corrections of order $\alpha_s^2$ terms not enhanced by  $\beta_0$, order $(\lqcd/m_b)^2$ terms proportional to derivatives of  $\delta(\shat)$, and higher order terms in both series. The result for the function $\gcut$ is shown in Fig.~\ref{figfinalplot} for three different values of $\mxcut$  as a function of $\qcut^2$. 

The uncertainties in the OPE prediction for $\gcut$ come from three separate sources: perturbative uncertainties from the unknown full two-loop result, uncertainties in the $b$ quark mass and uncertainties due to unknown matrix  elements of local operators at $O(1/m_b^3)$ in the OPE.   In the following I will consider each of these uncertainties separately as the
fractional  errors on $\gcut$.  The fractional uncertainty in $|V_{ub}|$ then is one half of the
resulting value. Starting with the perturbative uncertainties, we may estimate the error in the perturbation series in two ways: (a) as the
same size as the last term computed, the order $\alpha_s^2 \beta_0$ term, or (b) as the
change in the perturbation series by varying $\mu$ over some reasonable range.
These are illustrated in Fig.~\ref{figperturbative} (a) and (b), respectively. In
Fig.~\ref{figperturbative}(b) we vary the renormalization scale between
$\mu=4.7\,\gev$ and $\mu=m_b/3\sim 1.6\,\gev$, and plot the change in the
perturbative result.  For
a given set of $\qcut^2$ and $\mxcut$, we take the perturbative error to be the
larger of (a) and (b).
\EPSFIGURE{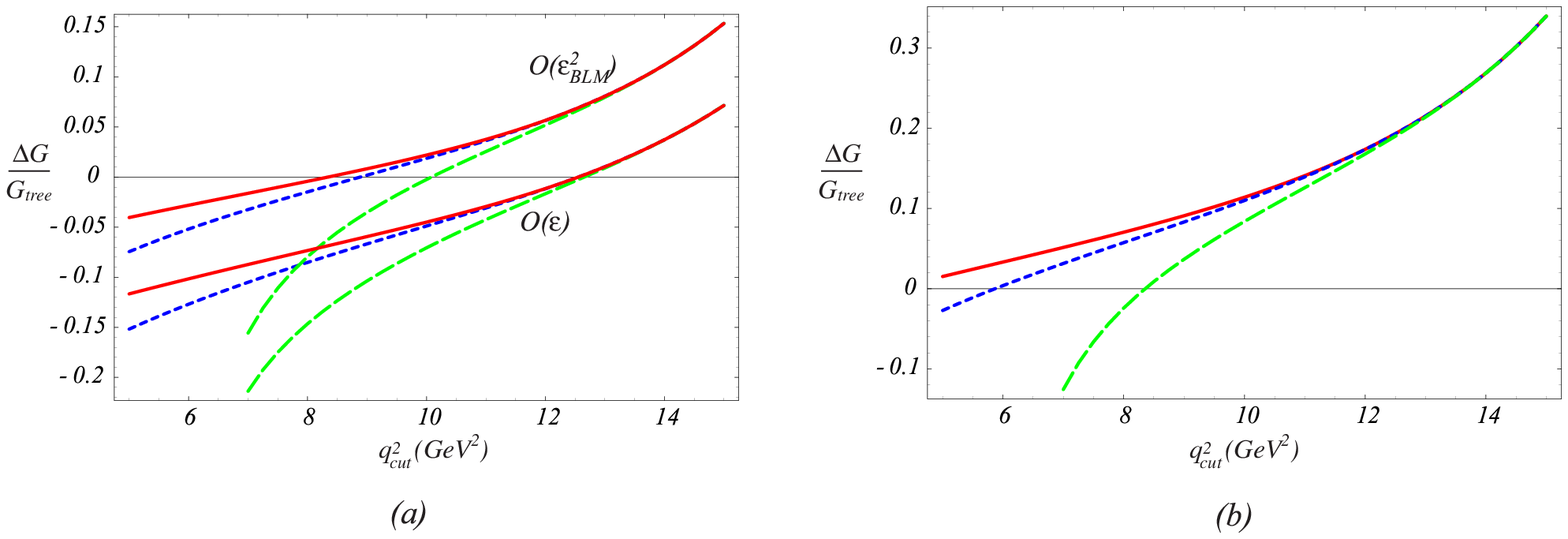,width=12cm}
{(a) The $O(\epsilon)$ and  $O(\epsblm)$ contributions to
$\gcut$
(normalized to the tree level result) for 
$\mxcut=1.86\,\gev$ (solid lines), $1.7\,\gev$ (short dashed lines) and
$1.5\,\gev$ (long dashed lines).  (b) Scale variation of the perturbative
corrections: The difference between the perturbative
corrections to $\gcut$, normalized to the tree level
result, for $\mu=4.7\,\gev$ and $\mu=1.6\,\gev$.\label{figperturbative}}

The partially integrated rate depends sensitively on the value of the $b$ quark
mass due both to the $m_b^5$ factor in $\gcut$ and the cut on $q^2$, as stressed
in \cite{neubertq2}.
Currently, the smallest error of the $1S$ mass is quoted from sum rules
\cite{sumrules1,benekesigner,hoang00}. Ref.~\cite{hoang00} obtains the value
$\mbups = 4.69\pm0.03\,\gev$ by fitting an optimized linear combination of
moments of the $e^+e^-\to b\,\bar b$ spectrum, which may underestimate the
theoretical error \cite{benekesigner}; the authors of \cite{benekesigner} cite a
 similar central value with a more conservative error of $\pm  0.08\,\gev$. In
Fig.~\ref{mass} we show the effects of a $\pm 80\,\mev$ and a $\pm 30\,\mev$
uncertainty in $\mbups$ on $\gcut$, using the central value $\mbups =
4.7\,\gev$. 
\EPSFIGURE{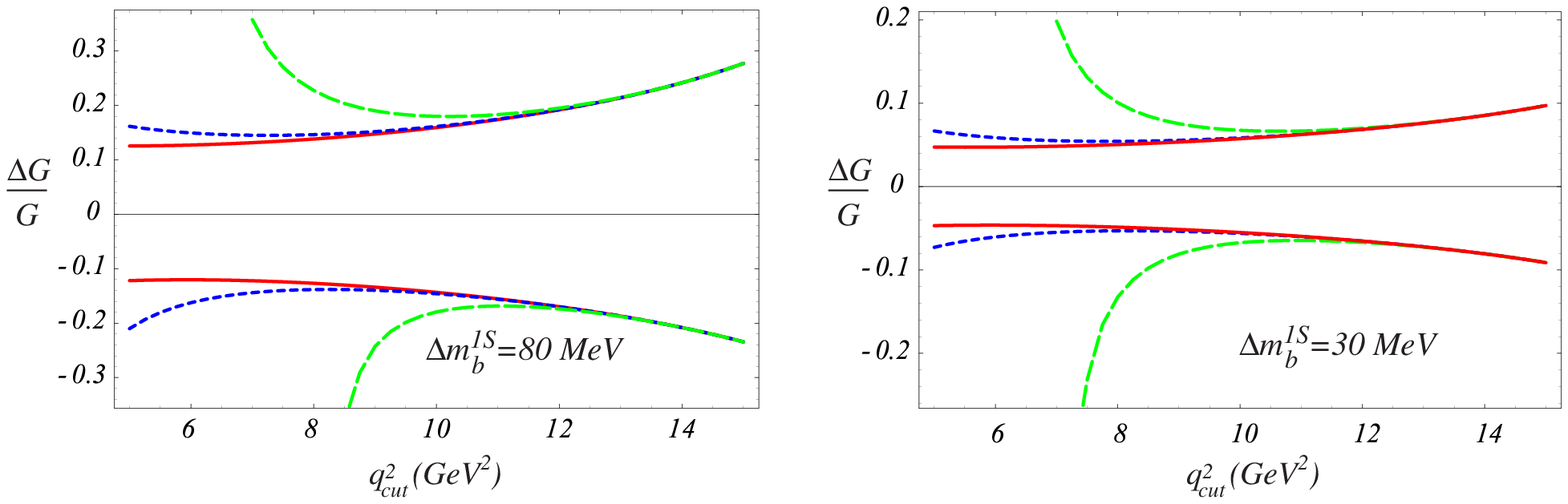,width=12cm}
{The fractional effect of a $\pm 80\,\mev$ and $\pm 30\,\mev$
uncertainty in $\mbups$ on $\gcut$ for $\mxcut=1.86\,\gev$ (solid line),
$1.7\,\gev$ (short dashed line) and $1.5\,\gev$ (long dashed line).\label{mass}}

Finally, there are uncertainties from unknown $1/m_b^3$ contributions to $\gcut$. We distinguish two contributions, the first originating from the local OPE \cite{mcubed} and the second from weak annihilation (WA) \cite{WA}. The latter terms vanish in the factorization limit, however they are enhanced by a factor of $16 \pi^2$. The two contributions are shown in Fig.~\ref{figmcubed}, assuming a 10\% violation of factorization, with the uncertainty from WA dominating. Since the WA contribute in general differently to charged and neutral $B$ decays, an indication of the size of the WA effect can be obtained by comparing the measurement of $\gcut$ from charged and neutral $B$'s. 
\DOUBLEFIGURE{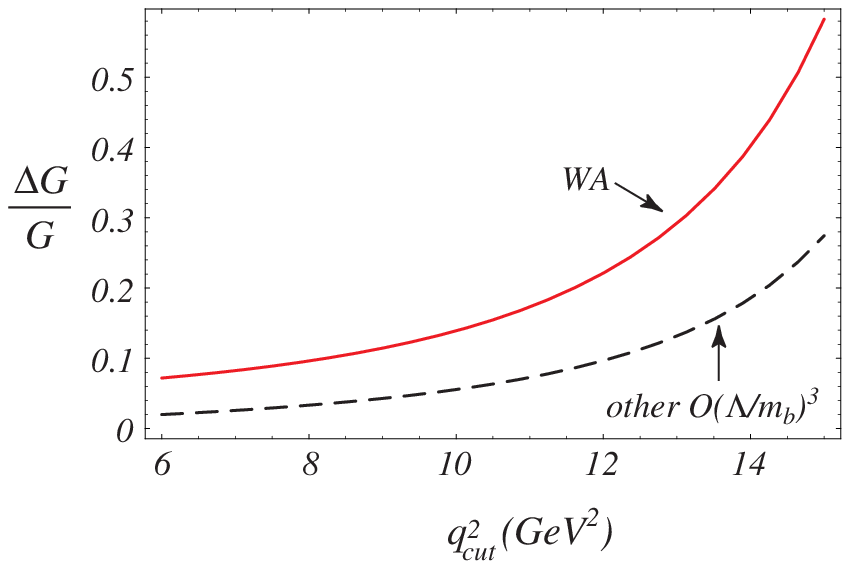,width=6cm}{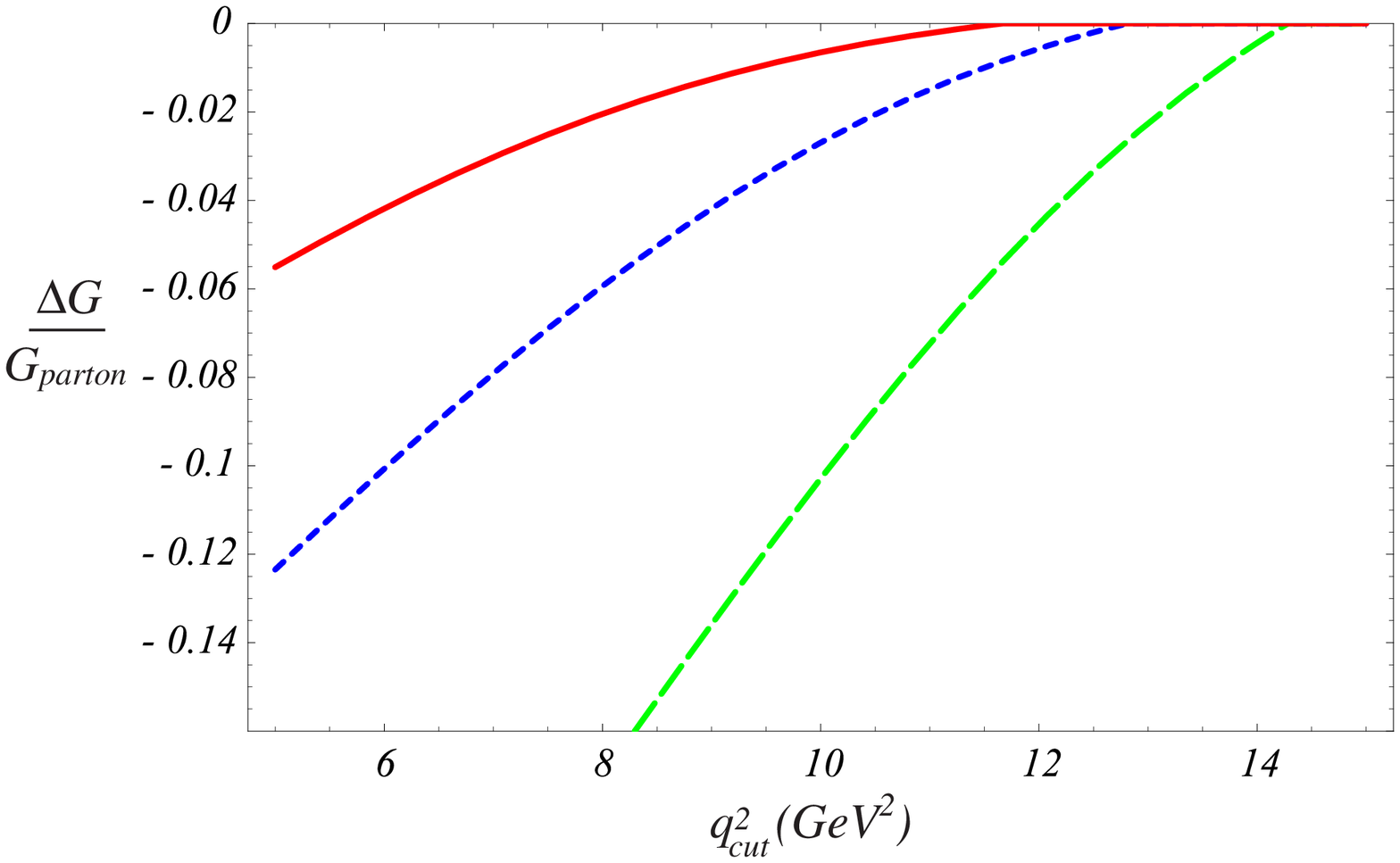,width=6cm}
{Estimate of the uncertainties due to dimension-six terms in the
OPE as a function of $\qcut^2$ from weak annihilation (WA) (solid line) and other operators
(dashed line).\label{figmcubed}}
{Effect of a model structure function on $\gcut$ as a function of
$\qcut^2$ for $\mxcut = 1.86\,\gev$ (solid line), $1.7\,\gev$ (short
dashed line) and $1.5\,\gev$ (long dashed line).\label{figSfig}}

Since the structure function becomes more important as the cut on $q^2$ is lowered, we still need to estimate the effect of the distribution function to  determine
how low $\qcut^2$ may be decreased.  To leading twist, this is obtained by smearing
the $b$ quark decay rate with the distribution function $f(k_+)$. The best way to determine $f(k_+)$ is from the $B\to X_s\gamma$ photon
spectrum, however in the absence of precise data, we will use a simple model to estimate the effects of the structure function.
In Fig.~\ref{figSfig} we plot in such a model the effect of the structure function
on $\gcut$ as a function of $\qcut^2$, for three different values of  $\mxcut$.

Using these results the strategy to extract $|V_{ub}|$ in a model independent way is: 
\begin{itemize}
\item make the cut on $m_X$ as large as possible, keeping the background from 
$b\to c$ small 
\item for a given cut on $m_X$, reduce the $q^2$ cut as low as possible, keeping the contribution from the $b$ quark structure function, as well as the perturbative uncertainties, small
\end{itemize}

\TABLE{
{\small
\begin{tabular}{c||c|c|c}
$\begin{array}{cc}
(\qcut^2,\mxcut)\\({\rm GeV^2},\, {\rm GeV}) \end{array}$  & $\begin{array}{c}{\rm Fraction}\\{\rm of} \,\,{\rm Events}\end{array}$ & ~$\Delta_{f}V_{\rm ub}$~ & $\begin{array}{cc}\Delta  V_{\rm ub}\\ {\footnotesize \pm80/30}\end{array}$  \\ \hline\hline
\multicolumn{1}{c}{Combined cuts} & \multicolumn{3}{c}{} \\ \hline
$(6,\,1.86)$	&  46\%  &  $ -2\%$  &
8\%/5\%  \\
$(8, 1.7)$ 	&  33\% &  $-3\%$&
9\%/6\% \\  \hline\hline
\multicolumn{1}{c}{Pure $q^2$ cut} & \multicolumn{3}{c}{} \\ \hline
~$(11.6, \,1.86)$~ & 17\% & --\,--& 
~15\%/12\%
\end{tabular}
}
\caption{Fraction of events, effect of structure function and uncertainty on $|V_{ub}|$ for two different combination of cuts, in comparison with a pure $q^2$ cut. \label{tabresult}}}
In Table \ref{tabresult} we show the results for two different combinations of cuts. From this table one can see that the number of events included using the combined cuts is about a factor of three larger than for a pure $q^2$ cut and that the uncertainties on $|V_{ub}|$ can be reduced by a factor of two. 

In summary, combining a cut on $q^2$ with a cut on $m_X$ allows for a model independent determination of $|V_{ub}|$ using up to 45\% of the total $b \to u$ events with uncertainties at the 5-10\% level. This is a significant improvement over other methods presented to date, which either have strong model dependence or larger theoretical uncertainties.

\end{document}